\documentstyle[11pt]{article}
\begin{document}
\newcommand{\R}{\frac{C_2(R)}{T(R)}}
\newcommand{\M}{{\frac{1}{M}}}
\newcommand{\G}{\frac{C_2(G)}{T(R)}}
\newcommand{\partialslash}{\partial \!\!\! /}
\newcommand{\xslash}{x \!\!\! /}
\newcommand{\half}{\mbox{\small{$\frac{1}{2}$}}}
\renewcommand{\refname}{References.}
\title{The Nambu--Jona-Lasinio model at $O(1/N^2)$.}
\author{J.A. Gracey, \\ Department of Applied Mathematics and Theoretical
Physics, \\ University of Liverpool, \\ P.O. Box 147, \\ Liverpool, \\
L69 3BX, \\ United Kingdom.}
\date{}
\maketitle
\vspace{5cm}
\noindent
{\bf Abstract.} We write down the anomalous dimensions of the fields of the
Nambu--Jona-Lasinio model or chiral Gross Neveu model with a continuous
global chiral symmetry for the two cases $U(1)$ $\times$ $U(1)$ and $SU(M)$
$\times$ $SU(M)$ at $O(1/N^2)$ in a $1/N$ expansion.

\vspace{-16cm}
\hspace{10cm}
{\bf LTH-308}
\newpage
The Nambu--Jona-Lasinio, (NJL), model was introduced in \cite{1} as a theory
with a continuous global chiral symmetry which is broken dynamically. It has
also been studied in the context of hadronic physics following the early work
of \cite{A} and lately in, for example, \cite{2}, and chiral perturbation
theory \cite{3} since it corresponds to a low energy effective theory of the
strong interactions. Further, there has also been significant interest in
analysing models with four fermi interactions, like the NJL model, in other
contexts such as the standard model, \cite{B,4,5}. For instance, it was
initially pointed out in \cite{B} and subsequently studied in more detail in
\cite{4,5} that the Higgs boson of the standard model could be regarded as a
composite field built out of fermions in much the same way that the $\sigma$
and $\pi$ fields of the two dimensional chiral Gross Neveu, (CGN), model are
bound states of the fundamental fermions of the model, \cite{6}.

One of the most widely used techniques to examine the CGN or NJL models is the
large $N$ expansion where the number of fundamental fields $N$ is allowed to
become large, \cite{6}, and $1/N$ can therefore be used as a dimensionless
coupling constant. Consequently one can show, for example, that the models
are renormalizable in $2$ $\leq$ $d$ $<$ $4$ dimensions in this approach
\cite{7} and simultaneously deduce that they possess a rich structure, such as
dynamical symmetry breaking. Whilst the leading order large $N$ analysis is
well understood, it is important to go beyond the leading order to improve our
knowledge of the quantum structure. Recently, new techniques to achieve this
for theories with fermionic interactions were introduced in \cite{8} for the
$O(N)$ Gross Neveu model based on the critical point self consistency methods
to calculate critical exponents in the bosonic $O(N)$ $\sigma$ model,
\cite{9,10}. In this letter we present the results of the application of the
same techniques to the $4$-fermi models with continuous chiral symmetry by
writing down the anomalous dimensions of each of the fields in arbitrary
dimensions at $O(1/N^2)$. This is one order beyond any previous analysis and it
is necessary to have such higher order expressions in order to improve our
understanding of the areas mentioned above. A further motivation for such a
computation lies in the accurate independent evaluation of these quantities in
three dimensions in order to provide precise estimates to compare with recent
numerical simulations of $4$-fermi models, \cite{11}.

More concretely, the lagrangians of the models we have considered are,
\cite{6},
\begin{equation}
L ~=~ i \bar{\psi}^i\partialslash\psi^i + \sigma \bar{\psi}^i\psi^i
+ i\pi\bar{\psi}^i\gamma^5\psi^i - \frac{1}{2g^2}(\sigma^2 + \pi^2)
\end{equation}
for the NJL model with a global $U(1)$ $\times$ $U(1)$ chiral symmetry and
\begin{equation}
L ~=~ i \bar{\psi}^{iI}\partialslash\psi^{iI} + \sigma \bar{\psi}^{iI}
\psi^{iI} + i\pi^a\bar{\psi}^{iI}\gamma^5\lambda^a_{IJ}\psi^{iJ}
- \frac{1}{2g^2}({\sigma^{}}^2 + {\pi^a}^2)
\end{equation}
for the same model but with an $SU(M)$ $\times$ $SU(M)$ chiral symmetry. The
bosonic fields $\sigma$ and $\pi^a$ are auxiliary and $g$ is the perturbative
coupling constant. In (2) the generalized Pauli matrices, $\lambda^a_{IJ}$,
$1$ $\leq$ $a$ $\leq$ $(M^2 - 1)$, $1$ $\leq$ $I$ $\leq$ $M$ are normalized
to $\mbox{Tr}(\lambda^a\lambda^b)$ $=$ $4T(R)\delta^{ab}$ and we used the
conventions of \cite{13} and \cite{14}. Both (1) and (2) involve $N$-tuplets
of fermions $1$ $\leq$ $i$ $\leq$ $N$ and this $N$ will become our expansion
parameter. We also use the conventions of \cite{11} in defining the properties
of $\gamma^5$ as $\{\gamma^\mu,\gamma^5\}$ $=$ $0$, $\mbox{tr}(\gamma^5
\gamma^{\mu_1}\ldots\gamma^{\mu_n})$ $=$ $0$ and $(\gamma^5)^2$ $=$ $1$ with
$\mbox{tr}1$ $=$ $2$. We remark that our $\gamma^5$ conventions in
$d$-dimensions retain the anticommutativity property. This differs from the
definition given in \cite{C} which is one formulation used to perform
consistent renormalization calculations using dimensional regularization where
the spacetime dimension is changed to provide a way of handling infinities,
\cite{C,D}. There, \cite{C}, one loses Lorentz invariance in the full
$d$-dimensional space where $[\gamma^\mu,\gamma^5]$ $=$ $0$ when $\mu$ is not
an index in the physical spacetime. By contrast, the method of \cite{9} uses
propagators defined in arbitrary but {\em fixed} dimensions which are
consistent with Lorentz and conformal symmetry. Therefore it seems more
appropriate here to retain the anticommutativity of the $\gamma^5$ for a
Lorentz invariant formulation. (Indeed an alternative to \cite{C} for treating
$\gamma^5$ in dimensionally regularized calculations retains this condition,
\cite{D}.)

In \cite{8}, the model with $\pi$ $=$ $0$ was solved at $O(1/N^2)$ by the self
consistency approach of obtaining critical exponents and we have used the same
methods to treat (1) and (2). Briefly, this involves solving for critical
exponents at the $d$-dimensional fixed point of the field theory where the
model possesses an extra scaling or conformal symmetry. With this scaling
property one writes down the most general form the propagators of the fields
can take, consistent with Lorentz and conformal symmetry, where the powers of
the scaling form are related to the dimension and therefore the anomalous
dimension of the fields, \cite{11}. To obtain analytic expressions for these
critical exponents one represents the skeleton Dyson equations of various
Green's functions by the scaling forms and solves the resulting representation
of the Dyson equations for the unknown exponents order by order in $1/N$. The
power of the method is illustrated by the fact that the leading order results
are deduced algebraically and agree with previous work, whilst the new results
at the subsequent order are deduced with a minimal amount of effort, [10-12].

For (1), the asymptotic scaling forms of the respective propagators are,
in coordinate space, in the critical region,
\begin{equation}
\psi(x) ~\sim~ \frac{A\xslash}{(x^2)^\alpha} ~~,~~
\sigma(x) ~\sim~ \frac{B}{(x^2)^\beta} ~~,~~
\pi(x) ~\sim~ \frac{C}{(x^2)^\gamma}
\end{equation}
where $A$, $B$ and $C$ are amplitudes which are independent of $x$ and the
dimensions of the fields are defined as
\begin{equation}
\alpha ~=~ \mu + \half \eta ~~,~~ \beta ~=~ 1 - \eta - \chi_\sigma ~~,~~
\gamma ~=~ 1 - \eta - \chi_\pi
\end{equation}
where $d$ $=$ $2\mu$ is the spacetime dimension, $\eta$ is the fermion
anomalous dimension and $\chi_\sigma$ and $\chi_\pi$ are the anomalous
dimensions of the respective vertices. The anomalous dimensions are expanded
in powers of $1/N$ via, for example, $\eta$ $=$ $\sum_{i=1}^\infty \eta_i/N^i$.
By substituting the scaling forms (3) into the skeleton Dyson equations with
dressed propagators, which are illustrated in figs. 1-3, we were able to
determine $\eta_2$ using results obtained in \cite{11}. We found
\begin{eqnarray}
\eta_1 &=& - \, \frac{2\Gamma(2\mu-1)}{\Gamma(\mu-1)\Gamma(1-\mu)
\Gamma(\mu+1)\Gamma(\mu)} \\
\eta_2 &=& \eta^2_1 \left[ \Psi(\mu) ~+~ \frac{2}{\mu-1}
{}~+~ \frac{1}{2\mu} \right]
\end{eqnarray}
where $\Psi(\mu)$ $=$ $\psi(2\mu-1)$ $-$ $\psi(1)$ $+$ $\psi(2-\mu)$ $-$
$\psi(\mu)$ and $\psi(x)$ is the logarithmic derivative of the
$\Gamma$-function, as well as $\chi_{\sigma \, 1}$ $=$ $\chi_{\pi \, 1}$ $=$
$0$. Equation (6) represents the first $O(1/N^2)$ quantity to be determined
for (1).

The $O(1/N^2)$ corrections to the vertex anomalous dimensions were deduced by
following the analogous calculation for the $O(N)$ Gross Neveu model given in
\cite{14}. It involves studying the scaling properties of the $3$-point
functions at $O(1/N^2)$ using a method which developed the leading order work
of \cite{15,16} for the bosonic $\sigma$ model on $CP(N)$. Rather than
illustrate the large number of graphs which arise at $O(1/N^2)$, we have given
in figure 4 the basic structure of the distinct graphs which arise, though
the graphs with vertex counterterms have not been shown. The basic integrals
corresponding to each of the graphs have been given in \cite{14} and it was
therefore a straightforward exercise to manipulate the graphs which occur in
(1) to be proportional to integrals whose values are already known, \cite{14}.
For (1), at $O(1/N^2)$ we found that the degeneracy in the $(\sigma,\pi)$
sector was not lifted but unlike at leading order the fields now have a
non-zero anomalous dimension, which is a new feature, ie
\begin{equation}
\chi_{\sigma \, 2} ~=~ \chi_{\pi \, 2} ~=~
- \, \frac{\mu^2(4\mu^2-10\mu+7) \eta^2_1}{2(\mu-1)^3}
\end{equation}

We have also carried out the same calculation for the non-abelian case (2)
and we note the results obtained at leading order are
\begin{equation}
\eta_1 ~=~ \frac{\tilde{\eta}_1}{2} \left[\M+\R\right]
\end{equation}
where $\tilde{\eta}_1$ $=$ $-\,2\Gamma(2\mu-1)/[\Gamma(\mu+1)\Gamma(\mu)
\Gamma(1-\mu)\Gamma(\mu-1)]$ and
\begin{eqnarray}
\chi_{\sigma \, 1} &=& \frac{\mu\tilde{\eta}_1}{2(\mu-1)}
\left[\M-\R\right] \\
\chi_{\pi \, 1} &=& \frac{\mu\tilde{\eta}_1}{2(\mu-1)}
\left[ \R - \M - \frac{C_2(G)}{2T(R)} \right]
\end{eqnarray}
where $\lambda^a \lambda^a$ $=$ $4C_2(R) I$, $f^{acd} f^{bcd}$ $=$ $C_2(G)
\delta^{ab}$ and $C_2(R)$ $=$ $(M^2-1)/2M$, $C_2(G)$ $=$ $M$ for $SU(M)$,
\cite{12,13}. Several leading order exponents were calculated in \cite{11} for
$SU(2)$ $\times$ $SU(2)$ and our results are in agreement with them which
provides us with a partial check on our calculation. At next to leading order
the expressions are more involved compared to (6) and (7). We found
\begin{eqnarray}
\eta_2 &=& \frac{\tilde{\eta}^2_1}{4} \left[ \left(\M+\R\right)^2
\left(\Psi(\mu) + \frac{2}{\mu-1} + \frac{1}{2\mu} \right) \right. \\
&+& \left. \frac{\mu}{(\mu-1)} \left( \left( \M-\R\right)^2
+ \frac{C_2(G)C_2(R)}{2T^2(R)} \right)
\left( \Psi(\mu) + \frac{3}{2(\mu-1)} \right) \right] \nonumber
\end{eqnarray}
\begin{eqnarray}
\chi_{\sigma \, 2} &=& \frac{\mu\tilde{\eta}^2_1}{4(\mu-1)^2}
\left[ (2\mu-1)\left( \frac{1}{M^2} - \frac{C^2_2(R)}{T^2(R)} \right)
\left( \Psi(\mu) + \frac{1}{(\mu-1)}\right) \right. \nonumber \\
&+& \left. \frac{\mu C_2(R)C_2(G)}{2T^2(R)} \left( \Psi(\mu)
+ \frac{1}{(\mu-1)} \right) + \frac{3\mu}{2(\mu-1)} \left( \frac{1}{M}
- \frac{C_2(R)}{T(R)}\right)^2 \right. \nonumber \\
&+& \left. \frac{5\mu C_2(R)}{(\mu-1)T(R)} \left(\M - \R\right)
- \frac{2\mu}{(\mu-1)}\left(\frac{1}{M^2} - \frac{C^2_2(R)}{T^2(R)} \right)
\right. \nonumber \\
&+& \left. \frac{\mu}{2(\mu-1)}\left(\M+\R\right)^2
- \frac{\mu(2\mu^2-5\mu+4)}{(\mu-1)M} \left(\M + \frac{3C_2(R)}{T(R)} \right)
\right. \nonumber \\
&+& \left. \frac{\mu}{M}\left( 3(\mu-1)\Theta(\mu)
- \frac{(2\mu-3)}{(\mu-1)}\right) \left(\M - \R\right) \right]
\end{eqnarray}
and
\begin{eqnarray}
\chi_{\pi \, 2} &=& \frac{\mu\tilde{\eta}^2_1}{4(\mu-1)^2}
\left[ \Psi(\mu) + \frac{1}{(\mu-1)} \right] \nonumber \\
&& \times \left[(2\mu-1)\left(\M+\R\right)
\left(\R-\M-\frac{C_2(G)}{2T(R)}\right) \right. \nonumber \\
&&- \left. \frac{\mu C_2(G)}{2T(R)} \left(\R-\frac{2}{M}-\frac{C_2(G)}{2T(R)}
\right) \right] \nonumber \\
&+& \frac{3\mu^2\tilde{\eta}^2_1}{4(\mu-1)^3}
\left(\R-\M-\frac{C_2(G)}{2T(R)}\right)^2 \nonumber \\
&+& \frac{5\mu^2\tilde{\eta}^2_1}{16(\mu-1)^2M} \left[ 4 \left(\R-\M\right)
- \G \right] \nonumber \\
&+& \frac{2\mu^2\tilde{\eta}^2_1}{(\mu-1)^3} \left[ \frac{1}{M^2}
-\frac{C^2_2(R)}{T^2(R)} - \frac{C_2(G)}{2T(R)M} - \frac{C^2_2(G)}{8T^2(R)}
+ \frac{C_2(R)C_2(G)}{T^2(R)} \right. \nonumber \\
&+& \left. \frac{1}{16} \left( \left( \R+\M\right)^2 - \frac{C_2(G)}
{MT(R)} - \frac{3C_2(G)C_2(R)}{2T^2(R)} + \frac{C^2_2(G)}{2T^2(R)} \right)
\right. \nonumber \\
&-& \left. \frac{3}{16}\left( \M - \R + \frac{C_2(G)}{2T(R)} \right)^2
- \frac{(2\mu^2-5\mu+4)}{8} \left( \frac{3}{M^2}
+ \frac{C_2(R)}{T(R)M} \right. \right. \nonumber \\
&-& \left. \left. \frac{C_2(G)}{4T(R)}\left(\R-\frac{3}{M}\right)
+\left(\R-\M\right)^2 \right) \right. \nonumber \\
&-& \left. \frac{(\mu-1)^2}{8} \left( 3\Theta(\mu) - \frac{(2\mu-3)}{(\mu-1)^2}
\right) \left(\frac{1}{M^2} - \frac{C_2(R)}{MT(R)} + \frac{C_2(G)}{MT(R)}
\right. \right. \nonumber \\
&-& \left. \left. \left(\R-\M - \frac{C_2(G)}{4T(R)} \right)
\left( \R-\M - \frac{C_2(G)}{2T(R)} \right) \right) \right]
\end{eqnarray}
where $\Theta(\mu)$ $=$ $\psi^\prime(\mu)$ $-$ $\psi^\prime(1)$, and we have
used the results of \cite{12} in manipulating the $f^{abc}$ and $d^{abc}$
tensors which arise in the $3$-loop graphs of fig. 4. We have expressed our
results in as general a form as possible which allows one to check that each
expression does agree with the analogous results of \cite{8,14} and the $O(N)$
model.

We conclude with the observation that our results will prove to be extremely
useful in establishing which other models lie in the same universality class as
(1) and (2) since, for example, we have provided independent analytic
expressions which can now be expanded in powers of $\epsilon$ and compared with
$\epsilon$-expansions of critical exponents of other models deduced from the
corresponding perturbative renormalization group functions.

\vspace{1cm}
\noindent
{\bf Acknowledgement.} The author thanks Dr S.J. Hands for a brief
conversation.

\newpage
\noindent
{\Large {\bf Figure Captions.}}
\begin{description}
\item[Fig. 1.] Dressed skeleton Dyson equation for $\psi$.
\item[Fig. 2.] Dressed skeleton Dyson equation for $\sigma$.
\item[Fig. 3.] Dressed skeleton Dyson equation for $\pi$.
\item[Fig. 4.] Vertex corrections with dressed propagators.
\end{description}
\end{document}